# A novel configuration model for random graphs with given degree sequence

Xu Xinping(徐新平) [a)*], Liu Feng(刘峰)[a)]

[a]*Institute of Particle Physics, HuaZhong Normal University, Wuhan 430079, China*

[b] *Nano Science & Technology Research College, Huazhong Normal University, Wuhan 430079, China*

**Abstract:** *Recently, random graphs in which vertices are characterized by hidden variables controlling the establishment of edges between pairs of vertices have attracted much attention. Here, we present a specific realization of a class of random network models in which the connection probability between two vertices (i,j) is a specific function of degrees ki and kj. In the framework of the configuration model of random graphs, we find analytical expressions for the degree correlation and clustering as a function of the variance of the desired degree distribution. The expressions obtained are checked by means of numerical simulations. Possible applications of our model are discussed.*

**Keywords:** random graphs, configuration model, correlations

**1. Introduction**

Many natural and man-made complex systems can be fruitfully represented and studied in terms of networks or graphs. [1] A random graph is a collection of points, or vertices, with lines, or edges, connecting pairs of them at random. The study of random graphs has a long history. Starting with the influential work of Erdo¨s and Re´nyi in the 1950s and 1960s, [2−4] random graph theory has developed into one of the mainstays of modern discrete mathematics. However, the classical random graphs (or Poisson random graphs) have severe shortcomings as models of the real-world phenomena since it is a static model yielding small-world networks with a Poisson degree distribution. In fact, despite the network diversity, most of real networks share three prominent structural features: small average path length (APL), high clustering and scale-free (SF) degree distribution. Current attention has focused on proposing network models incorporating these features. To explain the small-world phenomenon (i.e., small APL), the idea of shortcuts has been proposed by Watts and Strogatz. [5] To reveal the emergence of scale-free degree distributions in real networks, the concept of evolving networks based on preferential attachment has been introduced by Barabási and Albert.[6] As a matter of fact, the models proposed to embody the fundamental characteristics can be roughly divided into two classes: static and evolving. The second class of causal networks encompasses, in particular, the famous Barabási-Albert (BA) model, whereas the configuration model[7-9] and the large group of networks with hidden variables [10–12] belong to the first class of static networks.

Random graph model in which vertices are characterized by hidden variables controlling the establishment of edges between pairs of vertices has been recently proposed by Boguñá and Pastor-Satorras.[11] This general model considers graphs in which each vertex has assigned a hidden variable, randomly drawn from a fixed probability distribution. Edges are assigned to pairs of vertices with a given connection probability, depending on the values of the hidden variables at the edge end points. Another alternative approach to generate random graphs is the configuration model, [15] which is a construction model that has the degree distribution as an input but is random in all other respects. Very recently, Fronczak and Hołyst have show that configuration model with expected degree sequence is equivalent to random graph model with hidden variables. [13] Here, we adopt the methodology of configuration model with given degree sequence. In configuration model, the total number of nodes is fixed to $N$ and degrees of all nodes $i$=1,2, . . . ,$N$ create a specific degree sequence $\{k_i\}$. Until now, nothing has been said about the connections between nodes. As a rule, random graphs with a given degree sequence are constructed in the following way: first, attach to each node $i$ a number $k_i$ of *stubs* (ends of edges); next, choose pairs of these stubs uniformly at random and join them together to make complete edges. The configuration model is defined to be an ensemble of graphs with desired degree sequences and each having equal weight.

A typical example of the configuration model has recently been proposed by Chung and Lu.[14] In their model,

each vertex $i$ is assigned a desired degree $k_i$ chosen from a certain distribution of interest, and then $m = \frac{1}{2}\sum_i k_i$ edges are placed between vertex pairs $(i, j)$ with probability $p_{ij} = \frac{k_i k_j}{2m}$.[15] This model has the features that the expected degree sequence of the ensemble of networks is equal to the desired degree sequence while degree sequence of the single network is not in general precisely equal to the desired degree sequence. Recently, Xu et al have find that the model of Chung and Lu is equivalent to maximum-entropy networks in the case of sparse limit.[16] More features of the Chung Lu model have been widely studied by a number of authors.[17,18]

In this paper, we introduce a novel configuration model for random graphs by modifying the vertex pair connection probability of the Chung Lu model. We present a parrallel work with this specific implementation of connection function in the framework of configuration model with given degree sequences. As we known, the connection function of the classical random graphs is a constant value while the influential model of Barabási Albert possesses a connection function as the model of Chung and Lu, and much effort has been devoted to the solvable properties for random graphs with the two types of connection function. Since much previous work has been concentrated on static random graphs with the two types of connection function, our specific connection function may enrich the class of random graphs models. The whole text is organized as follows. Section 2 gives a general description for our model. In section 3, we find analytical expressions for the degree correlation and clustering of our model and compare these properties to that of the previous random graph models. In Section 4, we compare the simulation result with the analytical result. Conclusions and discussion are given in the last part, Sec. 5.

**2. The model**

We propose a configuration model for random graphs, in which the connection probability between vertex pairs $(i, j)$ is provided as a function of degrees ki and kj in the form

$$p_{ij} = \frac{k_i + k_j}{N} - p = \frac{k_i + k_j - z}{N} \tag{1}$$

where N is the network size, p is the average connection probability and z is the average degree.

Let's do some analysis to the connection probability of our model. Once given the desired degree sequence $\{ki\}$, the network size $N$, total number of edges $l = \frac{1}{2}\sum_i k_i$ and average edge connection probability $p = \frac{2l}{N^2}$ are fixed. After a simple algebra, one can find that the expected degree of vertex i is

$$\bar{k}_i = \sum_j p_{ij} = k_i + \frac{\sum_j k_j}{N} - \frac{\sum_j k_j}{N} = k_i \tag{2}$$

The expected degree is equal to the desired degree (or given degree). Therefore we can study the properties of our model derived by requiring the expected degree sequence of a graph ensemble to match a given set of degree sequence of a real-world network.

**3. Degree correlation and clustering of our model**

Let us use the connection probability (1) to calculate the ensemble mean of the degree correlation and clustering coefficient. Since the expected degree is equal to the desired degree (or given degree), we use the expected degrees to implement the calculations. The degree correlation can be quantified by calculating the mean degree of the network neighbors of a vertex as a function of the degree $k$ of that vertex.[1] The mean sum of the degrees of the neighbors of a vertex $i$, which is denoted by $\bar{K}_i^{nn}$, can be given by

$$\overline{K}_i^{nn} = \sum_j p_{ij}\overline{k}_j = \sum_j (\frac{\overline{k}_i + \overline{k}_j}{N} - p)\overline{k}_j = \overline{k}_i \frac{\sum_j \overline{k}_j}{N} + \frac{\sum_j \overline{k}_j^2}{N} - p\sum_j \overline{k}_j .$$

$$= pN\overline{k}_i + \frac{N\sum_j \overline{k}_j^2 - (\sum_j \overline{k}_j)^2}{N^2} = pN\overline{k}_i + <\overline{k}^2> - <\overline{k}>^2 \qquad (3)$$

Thus the mean degree of a neighbor of vertex $i$ is equal to

$$\overline{k}_i^{nn} = \frac{\overline{K}_i^{nn}}{\overline{k}_i} = pN + \frac{Q}{\overline{k}_i} = z + \frac{Q}{\overline{k}_i} \qquad (4)$$

where $z = \frac{\sum_i \overline{k}_i}{N} = pN$ is the average expected degree and Q is variance of the expected degree distribution.

Throughout this paper, we denote the expected values or ensemble mean by a bar (e.g., $\overline{k}$), and the average over network vertices by a bracket (e.g., $<\overline{k}>$). Eq. (4) is composed of an uncorrelated item $z$ and a correlated item $Q/\overline{k}_i$, indicating that $(\overline{k}_i^{nn} - z)$ versus $\overline{k}_i$ follows a power law with exponent -1. Thus our model belongs to the correlated random graph model, in contrast to uncorrelated classical random graph model and Chung Lu model.

Incidentally, the degree correlation coefficient defined by Newman is given by [19]

$$r = \frac{\sum_i \overline{k}_i \overline{K}_i^{nn} - (2\overline{l})^{-1}(\sum_i \overline{k}_i^2)^2}{\sum_i \overline{k}_i^3 - (2\overline{l})^{-1}(\sum_i \overline{k}_i^2)^2} = -\frac{(<\overline{k}^2> - <\overline{k}>^2)^2}{<\overline{k}><\overline{k}^3> - <\overline{k}^2>^2} = -\frac{Q^2}{<\overline{k}><\overline{k}^3> - <\overline{k}^2>^2}. \qquad (5)$$

Thus our model has negative correlation and shows a disassortative mixing structure. Since a large number of real networks seem to exhibit negative degree correlation and there is no correlation in the classical random graph model and Chung Lu model, our model may be more close to real networks.

Next, we derive analytical formula for the clustering coefficient. Before we get the clustering coefficient of a certain vertex $i$, we need to calculate the ensemble mean of edges among its nearest neighbors. Consider two arbitrary vertices $m$ and $n$. The probability that these two vertices connecting to vertex $i$ simultaneously can be given by $p_{im} p_{in}^{im}$, where $p_{in}^{im}$ is the connection probability between $i$ and $n$ on the condition that $i$ and $m$ are already connected. In the same way, we use $p_{mn}^{im,in}$ to denote the connection probability between $m$ and $n$, provided that they have a common vertex $i$. The average connections $\overline{E}_i$ between the nearest neighbors of vertex $i$, can be formally computed as the probability that vertex $i$ is connected to vertices $m$ and $n$, and that those two vertices are at the same time joined by an edge, summing over all the possible vertex pairs $m$ and $n$, and divided by 2. Thus, we can write the clustering coefficient of vertex $i$ as

$$\overline{C}_i = \frac{\overline{E}_i}{\overline{k}_i(\overline{k}_i - 1)/2} = \frac{2\overline{E}_i}{\overline{k}_i(\overline{k}_i - 1)} = \frac{\sum_{m,n} p_{im} p_{in}^{im} p_{mn}^{im,in}}{\overline{k}_i(\overline{k}_i - 1)} \qquad (6)$$

In our ensemble model, each link is present or absent independently of all others, therefore the conditional probability can be simplified as $p_{in}^{im} = p_{in}$, $p_{mn}^{im,in} = p_{mn}$. Substituting these formulas into (6), we get

$$\overline{C}_i \approx \frac{\sum_{m,n} p_{im} p_{in} p_{mn}}{\overline{k}_i^2} = p + \frac{2Q}{N\overline{k}_i} \tag{7}$$

Eq. (7) gives the relation between $\overline{C}_i$ and $\overline{k}_i$, which we call the clustering spectrum or degree-dependent clustering. [20-22] Eq. (7) indicates that $(\overline{C}_i - p)$ versus $\overline{k}_i$ also follows a power law with exponent -1. Considering that the classical random graph and Chung Lu model provide an insufficient description for the nontrivial clustering of real networks, we believe our model is more realistic. Consequently, the average clustering coefficient of the whole network is

$$C = p + \frac{2Q}{N^2} \sum_i \frac{1}{\overline{k}_i} = p + \frac{2Q}{N} <\frac{1}{\overline{k}}> \tag{8}$$

Combining Eq.(8) and (11), we get

$$\overline{C}_i = \frac{2\overline{k}_i^{nn}}{N} - p, \tag{9}$$

Therefore, the clustering coefficient $\overline{C}_i$ increases linearly with $\overline{k}_i^{nn}$, implying that degree-degree correlation leads to a degree-dependent local clustering. Previously, this feature was observed in a number of real networks. [23-25] Here we demonstrate that this dependence is a direct consequence of degree-degree correlation.

**4. Comparison with numerical simulation**

In order to test the analytical predictions, we do computer simulation in the following steps. First, we choose an advisable degree sequence from a certain distribution of interest as an input of the desired degree sequence. Second, we use the connection probability (1) to generate edges and finally the whole network as a realization in the ensemble. Third, repeat step 2, creating an ensemble of single networks. Fourth, calculate the corresponding variables including degree correlation and clustering for each single network, and get the average values of the variables in the ensemble networks. In order to minimize the statistical fluctuations, we average the physical quantities over a large number of realizations. Considering there is one link between any vertex pairs at most, the connection probability should satisfy $0 \leq P_{ij} \leq 1$ for any vertex pair $(i,j)$, the case of $p_{ij} < 0$ or $p_{ij} > 1$ can be avoided.

Since for most real networks, the degree distributions follow power-law $P(k) = ck^{-g}$ with $2 < g < 3$, we choose desired degree sequence from a power law with $g = 2.5$ to perform our numerical simulation. We expect other type of degree sequence should give qualitatively the same behavior. The main results are shown in Figs.1-3. Consider first Fig. 1, which is a plot of the mean degree $\overline{k}^{nn} - z$ as a function of $\overline{k}$. As the figure shows, $\overline{k}^{nn} - z$ versus $\overline{k}$ follows a power law with exponent -1, as expected. The same behavior is evident in Fig. 2 also, which shows $\overline{C} - p$ as a function of the mean degree $\overline{k}$ of that vertex. The straight line in Fig.1 and Fig.2 corresponds to a strict power law with exponent -1, as predicted by Eq. (4) and (7). In Fig.3, we plot the degree-dependent local clustering $\overline{C}$ versus the average nearest neighbor degree $k^{nn}$, which indicates a linear dependency between them. We have tried to alter the parameters of desired degree sequences, and similar behaviors or results are obtained.

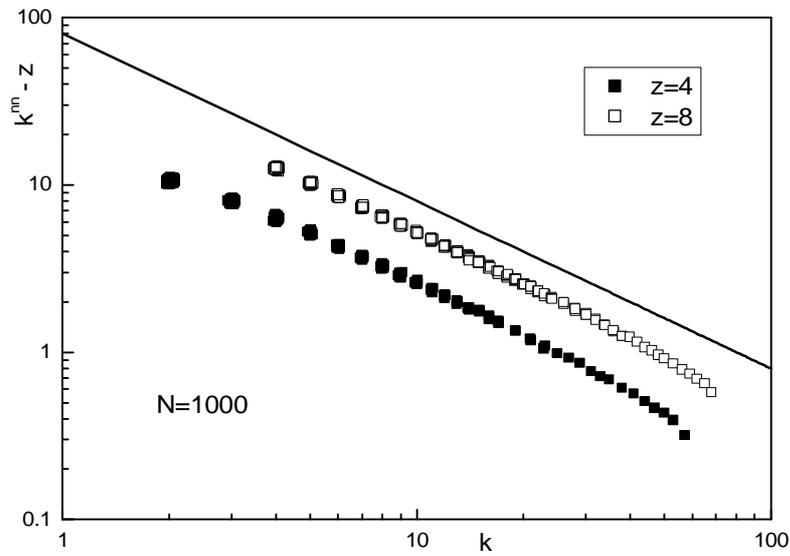

FIG.1. $\bar{k}^{nn} - z$ versus $\bar{k}$ in our model with $z = 4$ and $z = 8$. The scatters are the numerical results that are averaged over 5000 realizations for $N = 1000$. The continuous line corresponds with the analytical dependency $(k^{nn} - z) \sim k^{-1}$.

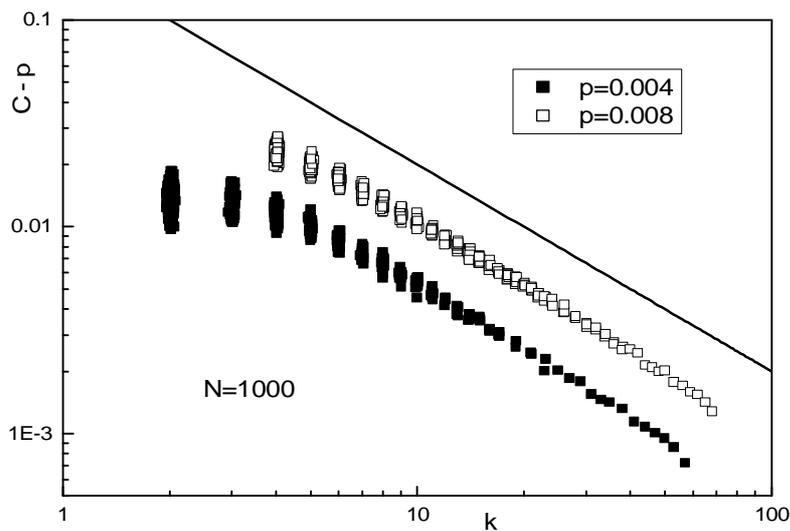

FIG.2. $\overline{C} - p$ versus $\bar{k}$ in our model with $p = 0.004$ and $p = 0.008$. The scatters are the numerical results that are averaged over 5000 realizations for $N = 1000$. The continuous line corresponds with the analytical dependency $(C - p) \sim k^{-1}$.

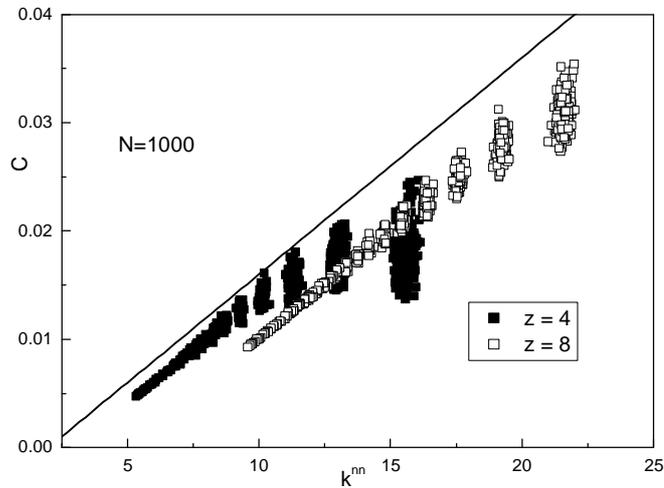

FIG.3. $\overline{C}$ versus $\overline{k}^{nn}$ in our model with $z = 4$ and $z = 8$. The scatters are the numerical results that are averaged over 5000 realizations for $N = 1000$. The continuous line has slope 0.002, as predicted by Eq.(9).

## 5. Conclusions and discussions

In this paper, we have presented a specific realization of a class of random network models in which the connection probability between two vertices (i,j) is a specific function of degrees ki and kj. We have derived the exact analytic formulas for the degree correlation and clustering of our model. It is found that the degree correlation and clustering are dependent on the variance of the given degree sequence. The analytical expressions are tested by numerical simulation.

Particularly, for the Poisson degree distribution, the second item of Eq.(8) is neglectable, the clustering coefficient of our model is $C \approx p$, Therefore, we can conclude that our model closes to classical random graph for the Poisson degree sequence. Furthermore, while the classical random graph model and Chung Lu model give a trivial description for the degree correlation, our model may provide a recommendable tool to unveil the origin of degree correlations observed in real networks. Many other properties of our model also require a further study, such as the position of the phase transition at which a giant component first forms, the mean component size, the size of the giant component if there is one, and the average vertex-vertex distance within a graph etc. Ultimately, we expect our model can shed much light on features or behaviors consistent with real-networks, and may take a useful step towards the study of random graphs.

**6. Acknowledgement** This work is supported by NSFC under projects 10375025, 10275027 and by the MOE under project CFKSTIP-704035.

*Corresponding author. E-mail address: xuxp@ihep.ac.cn